\documentclass[english,prd,twocolumn,superscriptaddress,nofootinbib,preprintnumbers]{revtex4}
\usepackage{graphicx}
\usepackage{amsmath,amsthm,amssymb}
\usepackage{bm}

\usepackage{amsfonts}
\usepackage{dcolumn}
\usepackage{hyperref}

\def\be{\begin{equation}}
\def\ee{\end{equation}}
\def\ba{\begin{eqnarray}}
\def\ea{\end{eqnarray}}
\def\bs{\begin{subequations}}
\def\es{\end{subequations}}

\usepackage{color}

\makeatother

\usepackage{babel}
\makeatother

\begin{document}

\title{Observational constraints on dark energy with a fast varying
equation of state}

\author{Antonio De Felice}

\affiliation{TPTP \& NEP, The Institute for Fundamental Study, Naresuan University,
Phitsanulok 65000, Thailand}
\affiliation{Thailand Center of Excellence in Physics, Ministry of Education,
Bangkok 10400, Thailand}

\author{Savvas Nesseris}

\affiliation{Departamento de Fisica Teorica and Instituto de Fisica
Teorica UAM/CSIC, Universidad Autonoma de Madrid, Cantoblanco,
E-28049 Madrid, Spain}

\author{Shinji Tsujikawa}

\affiliation{Department of Physics, Faculty of Science,
Tokyo University of Science,
1-3, Kagurazaka, Shinjuku-ku, Tokyo 162-8601, Japan}

\begin{abstract}

We place observational constraints on models with the late-time
cosmic acceleration based on a number of parametrizations
allowing fast transitions for the equation of state of dark energy.
In addition to the model of Linder and Huterer where
the dark energy equation of state $w$ monotonically grows or
decreases in time, we propose two new parametrizations
in which $w$ has an extremum.
We carry out the likelihood analysis with the three parametrizations
by using the observational data of supernovae type Ia,
cosmic microwave background, and baryon acoustic oscillations.
Although the transient cosmic acceleration models with
fast transitions can give rise to the total chi square smaller than that in the
$\Lambda$-Cold-Dark-Matter ($\Lambda$CDM) model,
these models are not favored over $\Lambda$CDM when one uses 
the Akaike information criterion which penalizes the extra degrees 
of freedom present in the parametrizations.

\end{abstract}

\date{\today}

\pacs{98.80.Cq, 95.30.Cq}

\maketitle

\section{Introduction}

The discovery of the late-time cosmic acceleration \cite{RP}
opened up a new research arena for cosmologists, astrophysicists,
and particle physicists.
{}From the viewpoint of particle physics the cosmological
constant naturally appears as a vacuum energy of quantum
fields, but its energy scale is usually very different from
the observed dark energy scale \cite{Weinberg}.
As an alternative to the cosmological constant,
dynamical dark energy models--such as quintessence \cite{quin},
k-essence \cite{kes}, $f(R)$ gravity \cite{fR}, $f(R,\mathcal{G})$ gravity \cite{fRG}, 
DGP braneworld \cite{DGP}, and Galileon \cite{Galileon}--have been proposed.
These models give rise to a time-varying equation of state $w(a)$
of dark energy, where $a$ is the scale factor
in the Friedmann-Lema\^{i}tre-Robertson-Walker (FLRW)
cosmological background \cite{review}.

At the background level it is possible to discriminate between
a host of dark energy models by confronting $w(a)$ with
the observations of Supernova type Ia (SN Ia), Cosmic
Microwave Background (CMB), and Baryon Acoustic
Oscillations (BAO).
For this purpose several different parametrizations of $w(a)$
have been proposed--which are mostly based on two
parameters $w_0$ and
$w_1$ \cite{Hut01,ChePo,Weller02,Linderpara,Jassal,Efstathiou}
(see Refs.~\cite{Hpara,Sahnireview} for the parametrizations 
of the Hubble parameter $H$ or the luminosity distance $D_L$
instead of $w$).
A well known example is the so-called
Chevalier-Polarski-Linder (CPL) parametrization
$w(a)=w_0+w_1(1-a)$ \cite{ChePo,Linderpara},
where $w_0$ is the value of $w$ today ($a=1$).
The 2-parameter parametrizations have been widely used
for constraining the property of dark
energy \cite{twopara,WMAP7,Suzuki}.

With two parameters one usually fixes $w_0$ and
the value of $w$ in the asymptotic past ($=w_p$).
In this case it is generally difficult to accommodate
the time $a_t$ and the width $\tau$ of the transition
in two asymptotic regimes.
Bassett {\it et al.} \cite{Bassett02} first proposed a 4-parameter
parametrization involving $a_t$ and $\tau$.
Corasaniti and Copeland \cite{Cora03} further developed
this issue and proposed a {\it kink} parametrization given by
$w(a)=w_0+(w_p-w_0) [(1+e^{a_t/\tau})(1-e^{(1-a)/\tau})]
[(1+e^{(a_t-a)/\tau})(1-e^{1/\tau})]^{-1}$.
This allows for quintessence models with tracker solutions
having a rapid transition \cite{Bare,Albre}, which is difficult to be
addressed by the 2-parameter parametrization.

Bassett {\it et al.} \cite{Bassett04} carried out the likelihood analysis
with the kink parametrization by using the Gold SN Ia
data \cite{Riess04} in 2004.
They found that the best-fit corresponds to the case in which
$w$ is nearly constant ($w \sim w_p= -0.41$) 
for the redshift $z$ larger than 0.1 and rapidly decreases 
toward $w_0 \sim -2.85$ for $z<0.1$.
This evolution of $w$ is outside the limits of the
two-parameter parametrizations, which implies
that two parameters are not generally sufficient to
implement such a rapid
transition\footnote{In some of quintessence and k-essence
models such as thawing and tracker models,
it is possible to derive the analytic forms of $w(a)$
approximately \cite{SchSen}-\cite{Chiba10}.
Apart from the tracker models with the
inverse power-law potential \cite{Chiba10},
the field equation of state usually contains more than
3 free parameters \cite{Dutta,Chiba09,Chiba09kes}.}.
Corasaniti {\it et al.} \cite{Corasaniti} also showed that 
the rapidly varying equation of state is consistent with 
the CMB and large-scale structure data
accumulated by 2004.

For the kink parametrization the Hubble parameter $H$
cannot be derived analytically in terms of a function
of $a$. Instead Linder and Huterer (LH) \cite{Linder05} proposed
the 4-parameter parametrization
$w(a)=w_f+(w_p-w_f)[1+(a/a_t)^{1/\tau}]^{-1}$,
which also allows a rapid transition (where $w_f$
is the value of $w$ in the asymptotic future).
In this case there exists an explicit integrated form of $H(a)$
with respect to $a$, so it is technically convenient.
Moreover this parametrization can accommodate tracker scaling
solutions ($w_p=0$) with the rapid decrease
of $w$ \cite{Bare,Cora03} and thawing quintessence
models ($w_p=-1$) with the fast growth
of $w$ \cite{Waga,Caldwell05}.

For both the kink and the LH parametrizations the dark energy
equation of state either increases or decreases
monotonically. Meanwhile there are some models in which 
$w$ has a minimum--such as
quintessence \cite{Bare}, $f(R)$ gravity \cite{fRm}, 
and coupled dark energy \cite{Marco} models.
In order to implement models in which $w$ has
an extremum, we propose two new parametrizations
given in Eqs.~(\ref{para2}) and (\ref{para3}) below.
These are based on four parameters $a_t$, $\tau$,
$w_p$, and $w_0$, which allow fast
transitions of $w$. Moreover, in both cases,
there exists analytic expression of the Hubble 
parameter\footnote{When the Hubble parameter is 
analytically available, one can also determine the 
$Om$ diagnostic introduced in Ref.~\cite{Om}.}.

In this paper we shall place constraints on the model
parameters of the LH parametrization
(\ref{para1}) as well as those of the parametrizations
(\ref{para2}) and (\ref{para3})
by using the recent observational data of SN Ia, CMB,
and BAO. In each model the five parameters
$a_t$, $\tau$, $w_p$, $w_0$ (or $w_f$), and
$\Omega_{m}^{(0)}$ (today's density parameter
of non-relativistic matter) are varied
in the likelihood analysis.
We also carry out the 4-parameter space analysis
by fixing $w_0$ with a number of different values
between $-1$ and $0$.
In order to accommodate the thawing-type models
with fast transitions, we shall further set
$w_p=-1$ and study the viability of the models 
(\ref{para2}) and (\ref{para3}) (including transient acceleration models)
in the 3-parameter space.
Note that the observational constraints on kink-like
parametrizations different from those mentioned above
(like those based on the deceleration parameter $q$)
have been studied by a number of
authors \cite{Ishida,Gui2010,Gio,Sha,Bruni}.

This paper is organized as follows.
In Sec.~\ref{parasec} we present the three
parametrizations of $w(a)$ as well as the
corresponding Hubble parameter $H(a)$.
In Sec.~\ref{datasec} we show the method of
our likelihood analysis to confront the models
with observations.
In Secs.~\ref{mo1sec}, \ref{mo2sec}, \ref{mo3sec}
we place observational constraints on the model
parameters of the parametrizations
(\ref{para1}), (\ref{para2}), and (\ref{para3}),
respectively. Sec.~\ref{consec} is devoted to conclusions.

\section{Parametrizations of dark energy}
\label{parasec}

We consider the flat FLRW background described
by the line element $ds^2=-dt^2+a^2 (t) d{\bm x}^2$,
where $t$ is cosmic time.
We take into account dark energy with the time-varying
equation of state $w(a)$ and non-relativistic matter
with the density parameter $\Omega_{m}^{(0)}$ today.
We assume that the dark energy density $\rho_{\rm DE}$
satisfies the continuity equation
\be
\dot{\rho}_{\rm DE}+3H (1+w) \rho_{\rm DE}=0\,,
\ee
where a dot represents a derivative with respect to $t$,
and $H=\dot{a}/a$ is the Hubble parameter.
This equation can be written in an integrated form
\be
\rho_{\rm DE}(a)=\rho_{\rm DE}^{(0)}
\exp \left[ \int_{a}^1 \frac{3}{\tilde{a}}
(1+w)\,d \tilde{a} \right]\,,
\ee
where $\rho_{\rm DE}^{(0)}$ is the dark energy density
today ($a=1$).
The energy density of non-relativistic matter
is given by $\rho_m(a)=\rho_m^{(0)}a^{-3}$, where
$\rho_m^{(0)}$ is its today's value.

The Friedmann equation gives
\be
3H^2=8\pi G (\rho_m+\rho_{\rm DE})\,,
\ee
where $G$ is the gravitational constant.
This can be written as
\be
\frac{H^2(a)}{H_0^2}=\Omega_m^{(0)} a^{-3}
+(1-\Omega_m^{(0)}) \exp \left[
\int_{a}^{1} \frac{3}{\tilde{a}} (1+w)
d\tilde{a} \right]\,,
\label{Hubble}
\ee
where $H_0$ is the present value of $H$,
$\Omega_m^{(0)}=8\pi G \rho_m^{(0)}/(3H_0^2)$,
and we used the fact that
$8\pi G \rho_{\rm DE}^{(0)}/(3H_0^2)=1-\Omega_{m}^{(0)}$.

Next, we study the parametrization of dark energy 
allowing fast evolution of $w$.
One of the examples is given by \cite{Linder05}
\be
w(a)=w_f+\frac{w_p-w_f}{1+(a/a_t)^{1/\tau}}
\qquad
({\rm Model}~1)\,,
\label{para1}
\ee
where $a_t~(>0)$ is the scale factor at the transition epoch,
and $\tau~(>0)$ characterizes the width of the transition.
In the asymptotic past ($a \to 0$) and future ($a \to \infty$)
one has $w \to w_p$ and $w \to w_f$, respectively.
For the parametrization (\ref{para1}) the r.h.s. of
Eq.~(\ref{Hubble}) is integrated to give
\ba
\frac{H^2(a)}{H_0^2} &=& \Omega_m^{(0)} a^{-3}
+(1-\Omega_m^{(0)})  \nonumber \\
&& \times a^{-3(1+w_p)} \left( \frac{a^{1/\tau}+a_t^{1/\tau}}
{1+a_t^{1/\tau}} \right)^{3\tau (w_p-w_f)}\,,
\label{Hubble1}
\ea
which is convenient in confronting the model with observations.

However, the parametrization (\ref{para1}) does not accommodate
the models in which $w$ has an extremum.
In order to address such cases, we propose
the following parametrization
\be
w(a)=w_p+(w_0-w_p)
\frac{a [1-(a/a_t)^{1/\tau}]}
{1-a_t^{-1/\tau}}
\qquad
({\rm Model}~2)\,,
\label{para2}
\ee
where $a_t>0$, $\tau>0$, and $w_p, w_0$ are the values
of $w$ in the asymptotic past and today, respectively.
For the parametrization (\ref{para2}) the Hubble
parameter can be expressed as
\be
\frac{H^2(a)}{H_0^2} = \Omega_m^{(0)} a^{-3}
+(1-\Omega_m^{(0)}) a^{-3(1+w_p)}
\exp \left[ f(a) \right]\,,
\label{Hubble2}
\ee
where
\ba
\hspace{-0.6cm}
f(a) &=& 3(w_0-w_p) \nonumber \\
\hspace{-0.6cm}
&\times&
\frac{1+(1-a_t^{-1/\tau})\tau
+a \{ [(a/a_t)^{1/\tau}-1]\tau-1 \}}
{(1-a_t^{-1/\tau})(1+\tau)}.
\ea
The equation of state (\ref{para2}) has an extremum at
\be
a_*=\left( \frac{\tau}{\tau+1} \right)^{\tau} a_t\,,
\label{astar}
\ee
with the value
\be
w(a_*)=w_p+\frac{(w_0-w_p) \tau^{\tau}(\tau+1)^{-\tau-1}a_t}
{1-a_t^{-1/\tau}}\,.
\ee
If $0<a_t<1$ and $w_p<w_0$, or, $a_t>1$ and $w_p>w_0$,
then $w$ has a minimum at $a=a_*$.
On the other hand, if $0<a_t<1$ and $w_p>w_0$,
or, $a_t>1$ and $w_p<w_0$,
$w$ has a maximum at $a=a_*$.

For the models characterized by $w_p>w_0$ with a minimum of $w$
at $0<a_*<1$ (such as quintessence models in Ref.~\cite{Bare}),
the transition redshift needs to satisfy the condition $a_t>1$.
{}From Eq.~(\ref{astar}) it follows that $a_t/e<a_*<a_t$ and
hence $a_*>1/e$.
This means that, for $w_p>w_0$, the parametrization (\ref{para2})
does not accommodate the case in which $w$ has a minimum
at low redshifts. In order to improve this shortcoming,
we shall also consider the following parametrization
\be
w(a)=w_p+(w_0-w_p)
\frac{a^{1/\tau} [1-(a/a_t)^{1/\tau}]}
{1-a_t^{-1/\tau}}
\qquad
({\rm Model}~3)\,,
\label{para3}
\ee
where $a_t>0$ and $\tau>0$.
Then $w$ has an extremum at
\be
a_*=\frac{a_t}{2^{\tau}}\,,
\label{minimo3}
\ee
with the value
\be
w(a_*)=w_p+\frac14 \frac{(w_0-w_p)
a_t^{1/\tau}}{1-a_t^{-1/\tau}}\,.
\ee
The equation of state has a minimum
either for $0<a_t<1$ and $w_p<w_0$, or,
for $a_t>1$ and $w_p>w_0$.
Since $a_* \to 0$ for $\tau \gg 1$,
we can cover the case of small $a_*$
even for $a_t>1$ and $w_p>w_0$.

The Hubble parameter corresponding to the
parametrization (\ref{para3}) is given by Eq.~(\ref{Hubble2}), 
where the function $f(a)$ is
\ba
\hspace{-0.3cm}
f(a) &=& 3(w_0-w_p) \tau \nonumber \\
\hspace{-0.3cm}
&\times&
\frac{2-a_t^{-1/\tau}+a^{1/\tau}[(a/a_t)^{1/\tau}-2]}
{2(1-a_t^{-1/\tau})}.
\ea

In the regime $0<a<1$ the parametrizations (\ref{para1}), (\ref{para2}),
and (\ref{para3}) can recover the CPL parametrization
$w(a)=w_0+w_1(1-a)$ in the limit that
$a_t \gg 1$ (with $\tau=1$ for Model 1 and Model 3).

\section{Data analysis}
\label{datasec}

In this section we explain the method employed to constrain
Models 1, 2, and 3 observationally.
In our analysis we use the three datasets:
1) the SN Ia (Constitution \cite{Constitution});
2) the CMB shift parameters (WMAP7) \cite{WMAP7};
3) the BAO (SDSS7) \cite{Percival}.
The flat Universe is assumed throughout the analysis.

In SN Ia observations the apparent magnitude $m(z)$
at peak brightness is related with the luminosity distance
$d_L(z)=(1+z) \int_{0}^z H^{-1} (\tilde{z})d \tilde{z}$
through $m(z)=M+5 \log_{10} (d_L(z)/10\,{\rm pc})$,
where $z=1/a-1$ is the redshift and $M$
is the absolute magnitude \cite{RP}.
We define the distance modulus
\be
\mu (z) \equiv m (z)-M=5 \log _{10}
[H_0 d_L(z)]+\mu_0\,,
\ee
where $\mu_0=42.38-5 \log_{10} h$ with
$h=H_0/[100\, {\rm km}\,{\rm sec}^{-1}\, {\rm Mpc}^{-1}]$.
The chi square associated with SN Ia observations is
given by
\be
\chi_{{\rm SN\, Ia}}^{2}=\sum_{i=1}^{N}
\frac{\mu_{{\rm obs}}
(z_{i})-\mu (z_{i})}{\sigma_{\mu,i}^{2}}\,,
\label{chisn}
\ee
where $N$ is the number of the SN Ia dataset,
$\mu_{{\rm obs}}(z_i)$ are the observed values
of the distance modulus, and $\sigma_{\mu,i}$ are
the errors on the data.
We employ the Constitution dataset with the total of
397 SN Ia in order to find the minimum of (\ref{chisn}) 
and the corresponding best-fit parameters.

The position of the CMB acoustic peaks can be quantified
by the following two parameters \cite{Bond}
\be
{\cal R}=\sqrt{\Omega_m^{(0)}} \int_0^{z_*}
\frac{dz}{H(z)/H_0}\,,\qquad
l_a=\frac{\pi d_a^{(c)} (z_*)}{r_s(z_*)}\,,
\ee
where $z_*$ is the redshift at the decoupling epoch,
$d_a^{(c)} (z_*)={\cal R}/[H_0 \sqrt{\Omega_m^{(0)}}]$ is
the comoving angular diameter distance
to the last scattering surface, and $r_s (z_*)$ is
the sound horizon defined by
\be
r_{s}(z_{*})=\int_{z_{*}}^{\infty}\frac{dz}{H(z)\,
\sqrt{3\{1+3\Omega_{b}^{(0)}/
[4\Omega_{\gamma}^{(0)}(1+z)]\}}}\,.
\ee
Here $\Omega_{b}^{(0)}$ and $\Omega_{\gamma}^{(0)}$
are the density parameters of baryons and photons,
respectively. For the redshift $z_*$ there exists
the following fitting formula \cite{Sugi}
\be
z_{*}=1048\,[1+0.00124(\Omega_{b}^{(0)} h^{2})^{-0.738}]\,
[1+g_{1}\,(\Omega_{m}^{(0)} h^{2})^{g_{2}}],
\ee
where $g_{1}=0.0783\,(\Omega_{b}^{(0)}h^{2})^{-0.238}
/[1+39.5\,(\Omega_{b}^{(0)} h^{2})^{0.763}]$ and
$g_{2}=0.560/[1+21.1\,(\Omega_{b}^{(0)} h^{2})^{1.81}]$.
The chi square for the WMAP7 measurement is
\be
\chi_{{\rm CMB}}^{2}={\bm X}_{\rm CMB}^T
\bm{C}_{{\rm CMB}}^{-1}{\bm X}_{\rm CMB}\,,
\ee
where
${\bm X}_{\rm CMB}^T=(l_{a}-302.09,{\cal R}-1.725,z_{*}-1091.3)$,
and the inverse covariance matrix is given by \cite{WMAP7}
\be
\bm{C}_{{\rm CMB}}^{-1}=\left(\begin{array}{ccc}
2.305 & 29.698 & -1.333\\
29.698 & 6825.27 & -113.18\\
-1.333 & -113.18 & 3.414
\end{array}\right)\,.
\ee

The BAO observations constrain the ratio
$r_{\rm BAO} (z) \equiv r_s (z_d)/D_V(z)$,
where $r_s(z_d)$ is the sound horizon at which the baryons are released from
the Compton drag of photons (denoted as the redshift $z_d$).
$D_V(z)$ is the effective BAO distance defined by
$D_V(z) \equiv [{d_a^{(c)}(z)}^2 z/H(z)]^{1/3}$ \cite{Eisen}, where
$d_a^{(c)}(z)=\int_0^{z} H^{-1} (\tilde{z})d \tilde{z}$.
For the redshift $z_d$ there is the following fitting formula \cite{EisenHu}
\be
z_{d}=\frac{1291\,(\Omega_{m}^{(0)} h^{2})^{0.251}}
{1+0.659\,(\Omega_{m}^{(0)} h^{2})^{0.828}}\,
[1+b_{1}\,(\Omega_{b}^{(0)} h^{2})^{b_{2}}]\,,
\ee
where $b_{1}=0.313\,(\Omega_{m}^{(0)} h^{2})^{-0.419}
[1+0.607\,(\Omega_{m}^{(0)} h^{2})^{0.674}]$
and $b_{2}=0.238(\Omega_{m}^{(0)} h^{2})^{0.223}$.
The chi square associated with the BAO measurement is given by
\be
\chi_{{\rm BAO1}}^{2}={\bm X}_{\rm BAO}^T
\bm{C}_{{\rm BAO}}^{-1}{\bm X}_{\rm BAO}\,,
\ee
where ${\bm X}_{\rm BAO}^T=(r_{\rm BAO}(0.2)-0.1905,
r_{\rm BAO}(0.35)-0.1097)$, and the inverse covariance
matrix is \cite{Percival}
\be
\bm{C}_{{\rm BAO}}^{-1}=\left(\begin{array}{cc}
30124 & -17227\\
-17227 & 86977
\end{array}\right)\,.
\ee

We also use the BAO data from the WiggleZ and 6dFGS surveys.
These data are given in terms of $A(z)$, where
its theoretical value is
\be
A_{\rm th}(z) \equiv
\frac{D_V(z) \sqrt{\Omega_m^{(0)} H_0^2}}{z}\,,
\ee
and the data are
$A_{\rm WiggleZ}(z=0.6)=0.452\pm0.018$ \cite{Blake:2011wn}
and $A_{\rm 6dFGS}(z=0.106)=0.526\pm0.028$ \cite{Beutler:2011hx}.
The chi-square is given by
\be
\chi_{{\rm BAO2}}^{2}=\sum_{i=1}^2\left(\frac{A(z_i)-A_{\rm th}(z_i)}
{\sigma_i}\right)^2\,.
\ee

Therefore, the total chi-square from the three datasets is
\be
\chi^2=\chi_{\rm SN\,Ia}^2+
\chi_{{\rm CMB}}^{2}+\chi_{{\rm BAO1}}^{2}+\chi_{{\rm BAO2}}^{2}\,.
\label{chis}
\ee
The best-fit corresponds to the model parameters
for which the $\chi^2$ is minimized.

\section{Observational constraints on Model 1}
\label{mo1sec}

We place observational constraints on Model 1
according to the method explained in Sec.~\ref{datasec}.

\begin{figure}
\includegraphics[height=3.3in,width=3.5in]{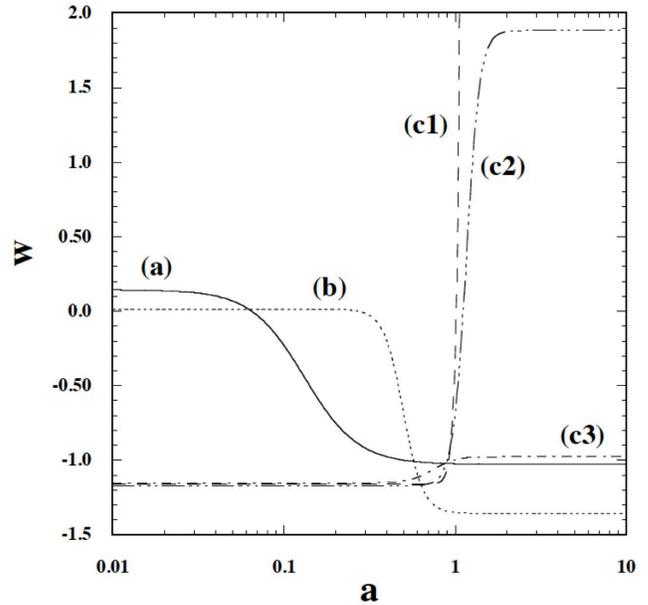}
\caption{\label{fig1}
The dark energy equation of state $w$ versus the scale factor $a$
for Model 1 with several different model parameters.
The line (a) corresponds to the best-fit case of Eq.~(\ref{5best}) derived
by varying the 5 parameters $w_p, w_f, a_t, \tau$, $\Omega_m^{(0)}$
in the likelihood analysis.
The line (b) shows the best-fit
derived with the priors $w_p \geq 0$ and $a_t \geq 0.5$.
The lines (c1), (c2), (c3) represent the best-fits
where the 4 parameters $w_p, a_t, \tau$, $\Omega_m^{(0)}$
are varied with the present value of the equation of state
fixed at $w_0=-1/3, -0.7, -1$, respectively.}
\end{figure}

We first vary the 5 parameters $w_p, w_f, a_t, \tau$,
and $\Omega_m^{(0)}$ in the likelihood analysis.
The priors on each parameter are set to be
$-10 \le w_p \le 10$, $-10 \le w_f \le 10$, $a_t>0$,
$\tau>0$, and $0.15<\Omega_m^{(0)}<0.4$.
We find that the best-fit parameters are
\ba
& & w_p=0.141866\,,\quad w_f=-1.02862\,,\quad
a_t=0.132023\,, \nonumber \\
& & \tau=0.360069\,,\quad
\Omega_m^{(0)}=0.290346\,,
\label{5best}
\ea
with $\chi^2=467.77$.
In Fig.~\ref{fig1} we plot the evolution of $w$ for the best-fit
case [line (a)]. Initially there is a period where
$w$ stays nearly constant ($w \simeq 0.14$),
which is followed by the decrease of
$w$ around the redshift $z$ larger than 10.
The dark energy equation of state crosses the cosmological
constant boundary ($w=-1$) around $z=1$ and it approaches
the asymptotic value $w_f=-1.028$.

If $w$ starts to evolve from the value larger than 0 in the
deep matter era, it is required that the transition
to the regime $w \approx -1$ occurs in the early
cosmological epoch (for $z$ larger than 1).
In fact, if we carry out the likelihood analysis
with the priors $w_p \geq 0$ and $a_t \geq 0.5$,
the best-fit model parameters are found to be
\ba
& & w_p=0.0120045\,,\quad w_f=-1.35856\,,\quad
~~a_t=0.5\,, \nonumber \\ \label{bf2}
& & \tau=0.130893\,,\quad
\Omega_m^{(0)}=0.300473\,,
\label{5best2}
\ea
with $\chi^2=505.983$. In this case, the bound on $a_t$ is saturated at
$a_t=0.5$ and the best-fit $\chi^2$ is much larger than
that corresponding to Eq.~(\ref{5best}).
Since $\tau \ll 1$, the transition from the regime $w \geq 0$
to the regime $w \approx -1$ occurs quite rapidly.
In Fig.~\ref{fig1} we compare the behavior of the two best-fits
in Eqs.~(\ref{5best}) and (\ref{5best2}).

The likelihood analysis of Bassett {\it et al.} \cite{Bassett04}
for the kink parametrization, with the SN Ia and CMB data accumulated 
by 2004, showed that the best fit corresponds to
the fast transition in low redshifts ($z<0.1$). However, inclusion of the BAO 
data as well as the more updated SN Ia and CMB data, 
seems to point towards a much earlier transition
from the regime $w \sim 0$ to the regime close to $w=-1$.

\begin{table}[t]
\begin{center}
\begin{tabular}{|c|c|c|c|c|c|}
\hline
$w_0$ & $w_p$ & $a_t$ & $\tau$ & $\Omega_m^{(0)}$ & $\chi^2$  \\
\hline
$0$ & $-1.04279$  & $1.20514$ & $0.00277414$ & $0.276338$ &
$470.825$ \\
\hline
$-1/3$ & $-1.16253$  & $1.31208$ & $0.0502325$ & $0.280804$ &
$471.530$ \\
\hline
$-0.5$ & $-1.17027$  & $1.30959$ & $0.0659688$ & $0.281219$ &
$470.996$ \\
\hline
$-0.6$ & $-1.17165$  & $1.26349$ & $0.0779737$ & $0.281295$ &
$470.707$ \\
\hline
$-0.7$ & $-1.17077$ & $1.16698$ & $0.0907055$ & $0.281248$ &
$470.456$ \\
\hline
$-0.8$ & $-1.16655$ & $1.18425$ & $0.116396$ & $0.280961$ &
$470.276$ \\
\hline
$-1$ & $-1.15384$ & $0.742678$ & $0.15533$ & $0.279799$ &
$470.387$ \\
\hline
\end{tabular}
\end{center}
\caption[model1chi]{
The best-fit model parameters (4 parameters in total)
and $\chi^2$ for Model 1 with several given values of $w_0$.
\label{model1chi} }
\end{table}

In order to study the possibility of the late-time transition
further, we also study the case in which the value of $w$ today ($=w_0$)
is fixed. Since $w_f=a_t^{1/\tau} [w_0 (1+a_t^{-1/\tau})-w_p]$,
the parametrization (\ref{para1}) can be expressed as
\be
w(a)=\frac{w_p+a^{1/\tau}[w_0 (1+a_t^{-1/\tau})-w_p]}
{1+(a/a_t)^{1/\tau}}\,.
\label{4para}
\ee
For several given values of $w_0$ we vary the 4 parameters
$w_p, a_t, \tau$, and $\Omega_m^{(0)}$ with the priors
$-10 \le w_p \le 10$, $a_t>0$, $\tau>0$,
and $0.15<\Omega_m^{(0)}<0.4$.
In Table \ref{model1chi} the best-fit model parameters
and the corresponding $\chi^2$ are shown
for $w_0=0, -1/3, -0.5, -0.6, -0.7, -0.8, -1$.
In Fig.~\ref{fig1} we also plot $w$ versus $a$
for several different best-fit cases ($w_0=-1/3, -0.7, -1$).

For the values of $w_0$ between $-1$ and 0, the initial
evolution of $w$ for each case shown in Table \ref{model1chi}
exhibits a common property.
The dark energy equation of state is nearly constant with
$w$ less than $-1$ during the deep matter era, which
is followed by the growth of $w$ in the low-redshift regime
($z \lesssim 1$).
The parameter $\tau$ tends to be smaller for larger $w_0$,
so that the transition becomes sharper.
This property can be confirmed by comparing the three
best-fit cases (c1)-(c3) in Fig.~\ref{fig1}.

In Table \ref{model1chi} we find that $\chi^2$ is more or less
similar for different choices of $w_0$ between $-1$ and $0$.
The best-fit $\Lambda$CDM model corresponds to
$\Omega_m^{(0)}=0.269431$
with $\chi^2=471.89$, whose $\chi^2$ is larger than
those given in Table \ref{model1chi}.
This implies that the transient cosmic acceleration models
with rapid transitions of $w$
are not excluded by the current observational data.

We need to caution, however, that the parametrization (\ref{4para})
with given $w_0$ has 4 free parameters to fit the models with the data,
while the $\Lambda$CDM model has only one free parameter
($\Omega_m^{(0)}$). In order to compare the models with
different number of free parameters, we employ
the Akaike Information Criterion (AIC) \cite{Liddle}.
The AIC is defined as
\be
{\rm AIC}=\chi_{\rm min}^2+2{\cal P}\,,
\ee
where $\chi_{\rm min}^2$ is the minimum value of $\chi^2$, and
${\cal P}$ is the number of free parameters for each model.
For smaller AIC the model is more favored.
If the difference of AIC between two models is in the
range $0<\Delta ({\rm AIC})<2$, the models are considered
to be equivalent.
On the other hand, if $\Delta ({\rm AIC})>2$,
one model is favored over another one.

The flat $\Lambda$CDM model corresponds to
${\rm AIC}=473.89$, whereas the transient acceleration
models in Table \ref{model1chi} give rise to larger values
of AIC (e.g., ${\rm AIC}=478.825$ for $w_0=0$).
The best-fit case (\ref{5best}) with 5 parameters
corresponds to ${\rm AIC}=477.77$.
According to the AIC, Model 1 with 5 or 4 free
parameters is not favored
over the $\Lambda$CDM model.

\section{Observational constraints on Model 2}
\label{mo2sec}

Let us proceed to observational constraints on Model 2.
We first vary the 5 parameters $w_p, w_0, a_t, \tau$,
and $\Omega_m^{(0)}$ in the likelihood analysis.
We set the priors on each parameter, as
$-10 \le w_p \le 10$, $-10 \le w_0 \le 10$, $a_t>0$,
$\tau>0$, and $0.15<\Omega_m^{(0)}<0.4$.
The best-fit parameters are found to be
\ba
& & w_p=-1.10237\,,\quad w_0=-0.906508\,,\quad
a_t=0.739325\,, \nonumber \\
& & \tau=0.505998\,,\quad \Omega_m^{(0)}=0.280583\,,
\label{5bestm2}
\ea
with $\chi^2=470.241$.
In this case $\chi^2$ is smaller than that in the $\Lambda$CDM
model, but it is larger than that in the best-fit case
(\ref{5best}) of Model 1.

\begin{figure}
\includegraphics[height=3.3in,width=3.5in]{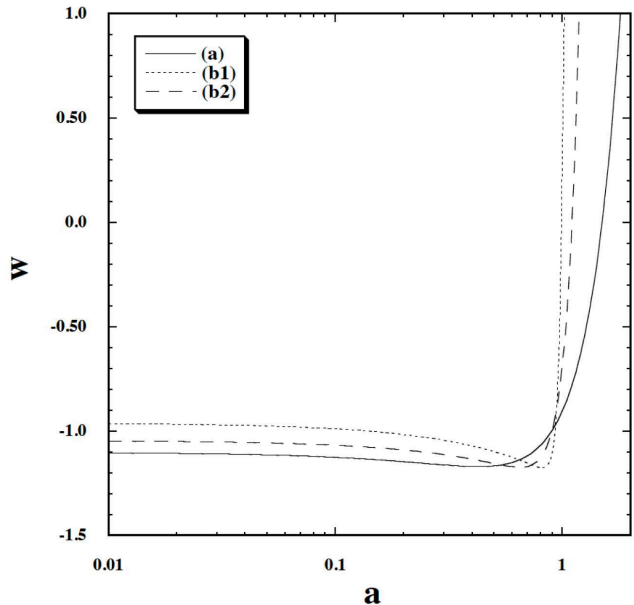}
\caption{\label{fig2}
The dark energy equation of state $w$ versus $a$ for Model 2.
The line (a) corresponds to the 5-parameter best-fit case
given in Eq.~(\ref{5bestm2}).
The lines (b1) and (b2) show the best-fits
where the 4 parameters $w_p, a_t, \tau$, $\Omega_m^{(0)}$
are varied with $w_0=0, -0.7$, respectively.}
\end{figure}

In Fig.~\ref{fig2} the evolution of $w$ for the parameters
(\ref{5bestm2}) is plotted as the solid line (a).
As we showed in Sec.~\ref{parasec}, $w$ has a minimum
at $a_*$ given in Eq.~(\ref{astar})
either for (i) $0<a_t<1$, $w_p<w_0$, or
(ii) $a_t>1$, $w_p>w_0$.
The best-fit model parameters (\ref{5bestm2}) correspond to
the case (i) with $a_*=0.43$.
The equation of state starts from a phantom value
$w_p=-1.10237$, which is followed by mild decrease of $w$.
For $a>a_*$ it starts to increase and reaches the present value
$w_0=-0.906508$.

The above behavior of $w$ is different from that
for the best-fit Model 1 with 5 parameters varied.
As we see in Fig.~\ref{fig1} the best-fit parameters
in Model 1 satisfy the condition $w_p>w_0$,
but in this case Model 2 gives rise to a minimum
only for $a_t>1$.
Since $a_*$ is limited in the range $a_t/e<a_*<a_t$ and
also $\tau$ is required to be large ($\tau \gg 1$)
to have small $a_*$, it becomes more difficult to fit $w$
with the observational data for $a_t>1$ and $w_p>w_0$.
If $w_p>w_0$ and $0<a_t<1$, $w$ has a maximum
at $a=a_*$. However, such cases are also difficult
to be compatible with the observational data.

By choosing several different values of $w_0$ ($=0, -1/3, -0.5, -0.7, -0.9$),
we also vary the 4 parameters $w_p, a_t, \tau, \Omega_m^{(0)}$
in the likelihood analysis. The priors are set to be $-10 \le w_p \le 10$,
$a_t >0$, $\tau>0$, and $0.15<\Omega_m^{(0)}<0.4$.
In Table.~\ref{model2chid} we show the best-fit parameters and $\chi^2$
for each $w_0$. In all cases we find that $0<a_t<1$ and $w_p<w_0$,
so that $w$ has a minimum at $0<a_*<1$.

%

\begin{table}[!t]
\begin{center}
\begin{tabular}{|c|c|c|c|c|c|}
\hline
$w_0$ & $w_p$ & $a_t$ & $\tau$ & $\Omega_m^{(0)}$ & $\chi^2$  \\
\hline
$0$ & $-0.96206$  & $0.93842$ & $0.04200$ & $0.27847$ & $473.130$ \\
\hline
$-1/3$ & $-1.05556$ & $0.90118$ & $0.06302$ & $0.28045$ & $471.638$ \\
\hline
$-0.5$ & $-1.14904$ & $0.80038$ & $0.07330$ & $0.28111$ & $471.001$ \\
\hline
$-0.7$ & $-1.04574$ & $0.87618$ & $0.14101$ & $0.28095$ & $470.468$ \\
\hline
$-0.9$ & $-0.97634$ & $0.92387$ & $0.77675$ & $0.28064$ & $470.246$ \\
\hline
\end{tabular}
\end{center}
\caption{The best-fit model parameters (4 parameters in total)
and $\chi^2$ for Model 2 with several given values of $w_0$.
\label{model2chid} }
\end{table}

\begin{table}[!t]
\begin{center}
\begin{tabular}{|c|c|c|c|c|c|}
\hline
$w_0$  & $w_p$ & $a_t$ & $\tau$ & $\Omega_m^{(0)}$ & $\chi^2$  \\
\hline
$0$    & $-1$ & $0.940663$ & $0.0319267$ & $0.277496$ & $472.984$ \\
\hline
$-1/3$ & $-1$ & $0.916762$ & $0.0689972$ & $0.280277$ & $471.702$ \\
\hline
$-0.5$ & $-1$ & $0.909223$ & $0.0938171$ & $0.280760$ & $471.074$ \\
\hline
$-0.7$ & $-1$ & $0.901454$ & $0.1517700$ & $0.280844$ & $470.476$ \\
\hline
$-0.9$ & $-1$ & $0.897160$ & $0.7207540$ & $0.280662$ & $470.245$ \\
\hline
\end{tabular}
\end{center}
\caption{The best-fit model parameters (3 parameters in total)
and $\chi^2$ for Model 2 with $w_p=-1$ and several given values of $w_0$.
\label{model2chid1}}
\end{table}

In Fig.~\ref{fig2} we show the variation of $w$ for the best-fit cases
with $w_0=0$ and $w_0=-0.7$ as the lines (b1) and (b2), respectively.
The growth of $w$ in the regime $a>a_*$ is sharper
for larger values of $w_0$.
This reflects the fact that, in Table \ref{model2chid},
$\tau$ gets smaller for $w_0$ increased.
We also note that the models with larger $w_0$ tend to be
disfavored because of the increase of $\chi^2$ seen
in Table \ref{model2chid}.

\begin{figure}
\includegraphics[height=3.2in,width=3.2in]{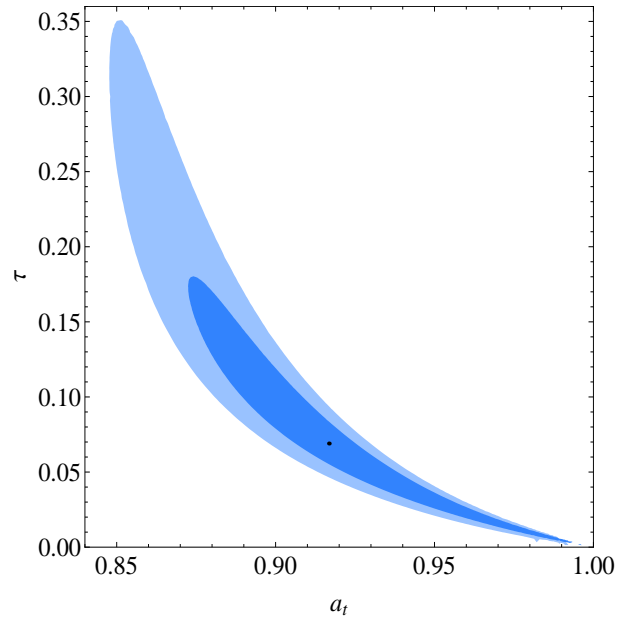}
\caption{\label{fig3}
The $1\sigma$ (inside) and $2\sigma$ (outside) likelihood contours
in the $(a_t, \tau)$ plane derived by varying the 2 parameters
$a_t$ and $\tau$ with $w_p=-1$, $w_0=-1/3$, and 
$\Omega_m^{(0)}=0.280277$ for Model 2.
The black point corresponds to the best-fit case.}
\end{figure}

In addition to $w_0$, we also fix $w_p$ to be $-1$ and vary the 3 parameters
$a_t, \tau, \Omega_m^{(0)}$. In Table \ref{model2chid1} we present
the best-fit model parameters for several different choices of $w_0~(>-1)$.
In all cases the transition scale factor is in the range $0<a_t<1$, 
so that $w$ has a minimum for $0<a_*<1$. 
In fact, the evolution of $w$ starting from $-1$ and having a minimum 
by today is present for dark energy models based on
$f(R)$ theories \cite{fRm}
(although $w$ does not continuously grow for $a>1$).
Table \ref{model2chid1} shows that, for larger $w_0$,
$\tau$ tends to be smaller, whereas $\chi^2$ gets larger.

In Fig.~\ref{fig3} we illustrate the $1\sigma$ and $2\sigma$ 
observational contours
in the $(a_t, \tau)$ plane for $w_p=-1$, $w_0=-1/3$, and 
$\Omega_m^{(0)}=0.280277$.
This is the marginal case in which the Universe
enters the phase of cosmic deceleration today.
The redshift and the width of the transition are
constrained to be $0.87<a_t<0.99$ and $0<\tau<0.18$ (68\% CL).
Unless the rapid transition occurs at the redshift close to today,
the model is not compatible with the observational data.
For larger $\tau$ the values of $w$ at $a=a_*$ start to deviate from
$-1$, so that those cases are more difficult to satisfy
observational constraints.

As mentioned in Sec.~\ref{mo1sec}, the AIC for the flat $\Lambda$CDM
model is ${\rm AIC}=473.89$.
For the 5-parameter best-fit case in Eq.~(\ref{5bestm2}) and for
the 4-parameter and 3-parameter best-fit cases given
in Tables \ref{model2chid} and \ref{model2chid1},
the AIC in each model is larger than
that in the $\Lambda$CDM model with 
the difference more than 2.
Hence the AIC criterion shows that the $\Lambda$CDM
is generally favored over Model 2.

\section{Observational constraints on Model 3}
\label{mo3sec}

Finally we proceed to observational constraints on Model 3.
When the 5 parameters $w_p, w_0, a_t, \tau,
\Omega_m^{(0)}$ are varied in the likelihood analysis,
we set the same priors as those given in
Model 2. The best-fit parameters  are found to be
\ba
& & w_p=-1.10733\,,\quad w_0=-0.897454\,,\quad
a_t=0.737871\,, \nonumber \\
& & \tau=0.652107\,,\quad \Omega_m^{(0)}=0.28053\,,
\label{5bestm3}
\ea
with $\chi^2=470.235$.

\begin{figure}
\includegraphics[height=3.3in,width=3.5in]{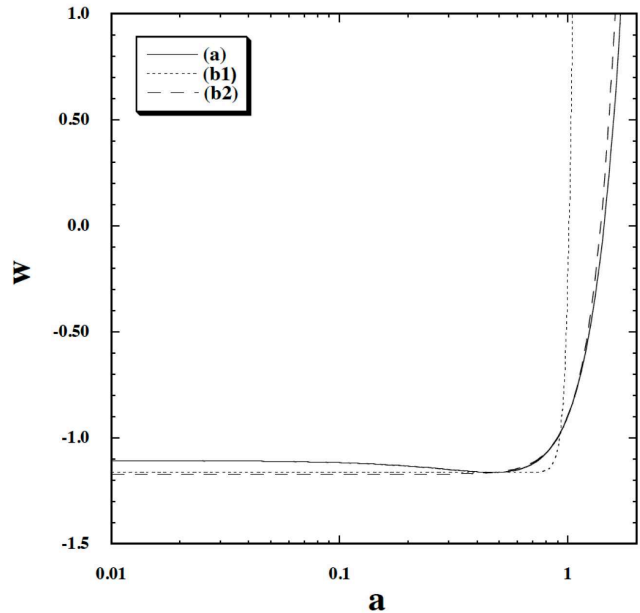}
\caption{\label{fig4}
The dark energy equation of state $w$ versus $a$ for Model 3.
The line (a) represents the 5-parameter best-fit case
given in Eq.~(\ref{5bestm3}), whereas the lines (b1) and (b2)
correspond to the best-fits derived by
varying the 4 parameters $w_p, a_t, \tau$, $\Omega_m^{(0)}$
with $w_0=-1/3, -0.9$, respectively.}
\end{figure}

As we see in Fig.~\ref{fig4}, the evolution of $w$ corresponding
to Eq.~(\ref{5bestm3}) is similar to that for the best-fit parameters
(\ref{5bestm2}) of Model 2.
Since $0<a_t<1$ and $w_p<w_0$ for the model
parameters (\ref{5bestm3}), $w$ has a minimum at $a_*=0.47$.
The difference between Models 2 and 3 is that
even for $a_t>1$ and $w_p>w_0$ the equation of state
for Model 3 can take minima with smaller values of
$a_*$ given in Eq.~(\ref{minimo3}).
However we find that the models with $a_t>1$ and $\tau \gg 1$
are disfavored because $w(a_*)$ deviates from $-1$.

For the 4-parameter parametrization with a number of different
values of $w_0$ ($=0, -1/3, -0.5, -0.7, -0.9$) we also
vary the parameters $w_p, a_t, \tau, \Omega_m^{(0)}$
with the same priors used for Model 2.
In Table \ref{model2chid31} we summarize the best-fit
parameters as well as the $\chi^2$ for each $w_0$.
For $w_0=-1/3, -0.5, -0.7$ one has $0<a_t<1$ and
$w_p<w_0$, in which cases $w$ has minima at
$0<a_*<1$. If $w_0>-0.5$, the growth of $w$
in the regime $a>a_*$ is very rapid
(see the line (b1) in Fig.~\ref{fig4}).

For the best-fit parameters corresponding to
$w_0=0, -0.9$ one has $a_t>1$ and $w_p<w_0$.
In those cases $w$ has maxima at $a_*$ larger than 1
and hence $w$ is a growing function for $a<1$.
Since $\tau$ is extremely small for $w_0=0$, the transition of
$w$ occurs almost like a step function.
However such an instant transition cannot be regarded
as a realistic model of dark energy.
For $w_0=-0.9$, $w$ has a maximum ($w (a_*)=5.5$)
at $a_*=2.4$. In this case the evolution
of $w$ is not very different from that for the best-fit case
(\ref{5bestm3}), apart from the fact that for $w_0=-0.9$
the equation of state is a growing function
in the regime $a<1$.

\begin{table}[!t]
\begin{center}
\begin{tabular}{|c|c|c|c|c|c|}
\hline
$w_0$ & $w_p$ & $a_t$ & $\tau$ & $\Omega_m^{(0)}$ & $\chi^2$  \\
\hline
$0$ & $-1.04278$  & $3.86853$ & $5.86338 \times 10^{-13}$ & $0.27634$ & $470.825$ \\
\hline
$-1/3$ & $-1.16195$  & $0.68613$ & $0.10267$ & $0.28079$ & $471.533$ \\
\hline
$-0.5$ & $-1.16009$  & $0.78777$ & $0.15968$ & $0.28113$ & $471.001$ \\
\hline
$-0.7$ & $-1.06189$ & $0.86545$ & $0.35609$ & $0.28086$ & $470.468$ \\
\hline
$-0.9$ & $-1.17212$ & $2.75038$ & $0.22092$ & $0.28049$ & $470.234$ \\
\hline
\end{tabular}
\end{center}
\caption{The best-fit model parameters (4 parameters in total)
and $\chi^2$ for Model 3 with several given values of $w_0$.
\label{model2chid31} }
\end{table}

\begin{table}[!t]
\begin{center}
\begin{tabular}{|c|c|c|c|c|c|}
\hline
$w_0$  & $w_p$ & $a_t$ & $\tau$ & $\Omega_m^{(0)}$ & $\chi^2$  \\
\hline
$0$    & $-1$ & $0.918824$ & $0.185756$ & $0.277967$ & $474.087$ \\
\hline
$-1/3$ & $-1$ & $0.909047$ & $0.249165$ & $0.279820$ & $472.013$ \\
\hline
$-0.5$ & $-1$ & $0.904406$ & $0.296622$ & $0.280332$ & $471.194$ \\
\hline
$-0.7$ & $-1$ & $0.899265$ & $0.398049$ & $0.280590$ & $470.495$ \\
\hline
$-0.9$ & $-1$ & $0.901692$ & $0.788501$ & $0.280323$ & $470.242$ \\
\hline
\end{tabular}
\end{center}
\caption{The best-fit model parameters (3 parameters in total)
and $\chi^2$ for Model 3 with $w_p=-1$ and several given values of $w_0$.
\label{model2chid32}}
\end{table}

We also vary the 3 parameters $a_t, \tau, \Omega_m^{(0)}$ by fixing $w_p$
to be $-1$ for several different choices of $w_0~(>-1)$.
In Table \ref{model2chid32} we show the best-fit values as well as $\chi^2$
for each $w_0$. In most cases the transition redshifts are around $a_t=0.9$.
As we increase $w_0$ the parameter $\tau$ gets smaller,
so that the transition occurs more rapidly.
For larger $w_0$, $\chi^2$ tends to be larger.

\begin{figure}
\includegraphics[height=3.2in,width=3.2in]{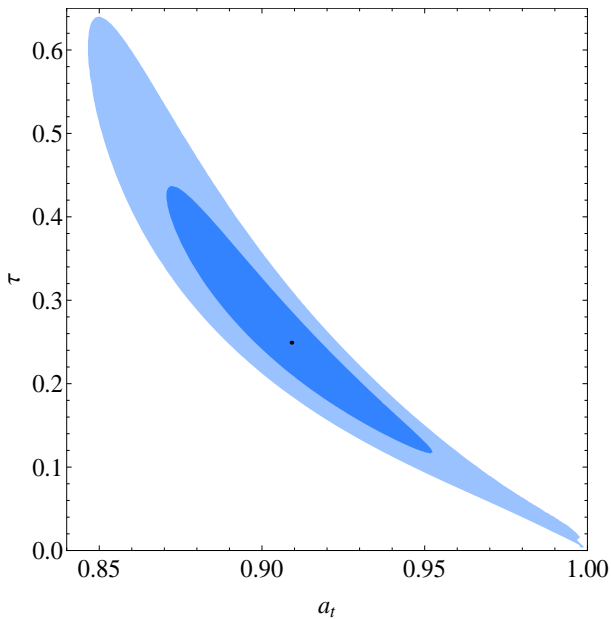}
\caption{\label{fig5}
The $1\sigma$ (inside) and $2\sigma$ (outside) likelihood contours
in the $(a_t, \tau)$ plane derived by varying the 2 parameters
$a_t$ and $\tau$ with $w_p=-1$, $w_0=-1/3$, and 
$\Omega_m^{(0)}=0.279820$ for Model 3.
The black point corresponds to the best-fit case.}
\end{figure}

In Fig.~\ref{fig5} we plot observational bounds in the $(a_t, \tau)$
plane for $w_p=-1$, $w_0=-1/3$, and $\Omega_m^{(0)}=0.279820$.
Comparing it to Fig.~\ref{fig3}, we find that
the larger values of $\tau$ can be allowed in Model 3.
This reflects the fact that in Model 3 the values
of $w(a_*)$ do not deviate from $-1$ significantly
for $\tau \lesssim 0.5$. The two parameters are constrained to be
$0.87<a_t<0.95$ and $0.12<\tau<0.44$ (68\% CL).

For all the best-fit cases discussed above, the AIC is larger than
that in the flat $\Lambda$CDM model.
Hence Model 3 is not favored over the $\Lambda$CDM
model according to the AIC.

\section{Conclusions}
\label{consec}

In this paper we placed observational constraints on the three
models allowing fast transitions of $w$, by using the data of
SN Ia, CMB shift parameters, and BAO.
Unlike the 2-parameter parametrization such as $w(a)=w_0+w_1(1-a)$,
the parametrizations (\ref{para1}), (\ref{para2}), and (\ref{para3}) have
two more parameters $a_t$ and $\tau$ by which the time and the
width of the transition can be accommodated.
In Model 1 the dark energy equation of state monotonically increases
or decreases in time, whereas in Models 2 and 3 $w$ has either a
minimum or a maximum depending on the values of $w_p$, $w_0$,
and $a_t$. For all these models the Hubble parameter $H$ is
analytically known in terms of functions of $a$.

When the 5 parameters $w_p, w_f, a_t, \tau, \Omega_m^{(0)}$
are varied in Model 1, the best-fit parameters are given by
Eq.~(\ref{5best}) with $\chi^2=467.77$.
This corresponds to the solid curve (a) in Fig.~\ref{fig1},
in which case the equation of state enters the regime $w \sim -1$
in the early cosmological epoch.
If we put the prior on the transition redshift as $a_t>0.5$,
$\chi^2$ becomes significantly larger than that without
a prior on $a_t$.
This means that the late-time transition ($a_t>0.5$) from
the regime $w \sim 0$ to the regime $w \sim -1$
is disfavored observationally.
If we vary the 4 parameters $w_p, a_t, \tau, \Omega_m^{(0)}$
with several different values of $w_0$ between $-1$ and $0$,
the parameter $\tau$ tends to be smaller for increasing $w_0$.
Although the $\chi^2$ in Model 1 with 5 or 4 parameters
can be smaller than that in the $\Lambda$CDM model, 
the AIC shows that Model 1
is not favored over the $\Lambda$CDM model.

The best-fit parameters for Model 2 corresponds to the case in
which $w$ starts from a phantom value $w_p=-1.10$, takes
a minimum $-1.17$ at $a_*=0.43$, and grows to the value
$w_0=-0.91$ by today. This is different from the evolution of
$w$ for the best-fit parameters of Model 1.
This difference mainly comes from the fact that, in the cases
$w_p>w_0$ for Model 2, $w$ has a minimum at $a_*$
given by Eq.~(\ref{astar}) only for $a_t>1$.
While Model 2 can accommodate the late-time transition having
a minimum of $w$, it is difficult to address the early sharp transition
with $w_p>w_0$.
The 4-parameter likelihood analysis for a number of fixed $w_0$
(between $-1$ and $0$) leads to similar best-fit evolution of $w$
to that for the 5-parameter best-fit case, 
with a faster transition for larger $w_0$.

In Model 3 the equation of state has an extremum at $a_*=a_t/2^{\tau}$,
which can be smaller that that for Model 2.
If $w_p>w_0$, however, the early transition of $w$ with a minimum
requires the condition $\tau \gg 1$.
This value of $\tau$ is too large to accommodate the early transition
compatible with observations, because the minimum value of $w$
tends to deviate from $-1$.
The 5-parameter likelihood analysis shows that the best-fit case
in Model 3 is similar to that in Model 2.
The likelihood results for 4 parameters ($w_0$ fixed) and
for 3 parameters ($w_0$ and $w_p$ fixed) also give rise
to the similar results to those found in Model 2.

In Models 2 and 3 the AIC is always larger than that in the
$\Lambda$CDM model with the difference more than 2.
Hence the models with the late-time fast transition to the
non-accelerating Universe are disfavored compared
to the $\Lambda$CDM model.
The joint data analysis based
on SN Ia, CMB, and BAO prefers the models in which
$w$ do not deviate significantly from $-1$
in the low-redshift regime.

Although the minimum value of $\chi^2$ in Model 1 with 5 parameters
is smaller than those in Models 2 and 3, the minima in Models 
2 and 3 are still inside the 1$\sigma$ region corresponding to Model 1.
Therefore we cannot prefer/exclude any parametrization 
with respect to any other one.

Recently it was shown that, in the framework of the CPL parametrization, 
the observational constraints on dark energy are sensitive to 
the presence of the cosmic curvature $\Omega_K^{(0)}$ \cite{Cardenas}.
They found that the CPL parametrization is not 
sufficiently flexible to model the rapidly varying equation of state
in low redshifts even for $\Omega_K^{(0)} \neq 0$
(see also Ref.~\cite{Sha}).
It will be of interest to study how the effect of
the cosmic curvature affects the observational constraints
on the models discussed in this paper. 
We leave this for future work.

\section*{ACKNOWLEDGEMENTS}
We thank Takeshi Chiba for useful discussions.
S.~N. acknowledges support from the Madrid Regional Government (CAM) 
under the program 
HEPHACOS S2009/ESP-1473-02.
S.~T. is supported by the Grant-in-Aid for
Scientific Research Fund of the Fund of the
JSPS No 30318802 and Scientific Research
on Innovative Areas (No.~21111006).
S.~T. thanks Reza Tavakol for warm hospitality during his stay
in University of Queen Mary. 

\end{document}